\documentstyle[preprint,epsf, aps]{revtex}

 {\par\noindent{\underline{Proof} \quad}}{\hfill$\Box$\bigskip}
 {\par\noindent{\underline{Proof} of the theorem\quad}}{\hfill$\Box$\bigskip}
 {\par\smallskip\noindent{\underline{{\it Remark}} \quad}}{\par\smallskip}
 {\par\smallskip\noindent{\underline{{\it Fact}} \quad}}{\par\smallskip}
 {\par\smallskip\noindent{\underline{{\it Example}} \quad}}{\par\smallskip}
 {\par\smallskip\noindent{{\it Assumotion} \quad}}{\par\smallskip}
 {\par\smallskip\noindent{{\it Condition} \quad}}{\par\smallskip}

\begin{document}
\title{Dynamics of Josephson junction systems in the computational subspace }
\author{{\bf Wang Xiangbin, Matsumoto Keiji, Fan Heng} \\
        Imai Quantum Computation and Information project, ERATO, Japan Sci. and Tech. Corp.\\
Daini Hongo White Bldg. 201, 5-28-3, Hongo, Bunkyo, Tokyo 113-0033, Japan\\
{\bf Y. Nakamura}\\NEC Fundamental Research
Lab., Tsukuba, Ibaraki 305-8051, Japan}

\maketitle
\begin{abstract} 
The quantum dynamics of the Josephson junction system in the computational subspace is investigated.
A scheme for the controlled not operation is given for two capasitively coupled SQUIDs. In this system,
there is no systematic error for the two qubit operation. 
For the inductively coupled SQUIDs, the effective Hamiltonian causes systematic errors in 
the computational subspace for the two qubit operation.
Using the purterbation
theory, we construct a more precise effective Hamiltonian. This new effective Hamiltonian reduces the
systematic error to the level much lower than the threshold of the fault resilent quantum computation.
\end{abstract}
\section{Introduction}
It has been shown that, a quantum computer, if available, can perform certain tasks much more
efficiently than a classical Turing machine\cite{loyd,bennett,di}. The realization of the basic
constitute of quantum computer, fault tolerate quantum logic gate, is a certral issue in the subject.
One can make a fault tolerate quantum computation through the quantum error correction. It has been 
pointed out that, if the error rate of each operation is smaller than certain threshold, the error 
correction method works for arbitrary large scale computation. The threshold is estimated to be $10^{-6}$.

Recently, it is proposed\cite{naka,rev,mak,shnirman} to implement the the quantum gate by super conducting interference device(SQUID).
The single qubit operation has been experimentally demonstrated in ref\cite{naka}.
This implementation has a unumber of advantages. One can address the single qubit( Josephson Junction) instead
of the bulk material in NMR. One can even control the interaction between the two qubit system through 
the external parameters( voltages and magnetic flux). However, strictly speaking, the Josephson junction
is not a two level system. Only when we choose certain specific parameters we can  approximately  have
a two level computational subspace. In general there is a small transition probability 
between the computational spapce and the outside space, this causes the leakage\cite{leak}. 
The leakage error can be minimized by the specifically designed farbrication\cite{leak}.
Also, one may detect and correct the leakage error by taking the meassurement on the space outside
computational space.
Besides the leakage, the approximate effective Hamiltonian causes another type of error, the
phase shift error. If the rate
of all accumulatable errors are lower than certain threshold\cite{laf}, a large reliable quantum
computation can be done, in principle. Because in such cases 
we can always take certain appropriate error correction procedure 
to reduce the error exponentially\cite{laf}. 
In this paper we analyse the phase shift error
caused by the approximate effective Hamiltonian. We give a new effective Hamiltonian through
the purtabation method. Using this new effective Hamiltonian, the phase shift error rate 
is lower than the quantum computation threshold\cite{laf}.  

\begin{figure}
\begin{center}
\epsffile{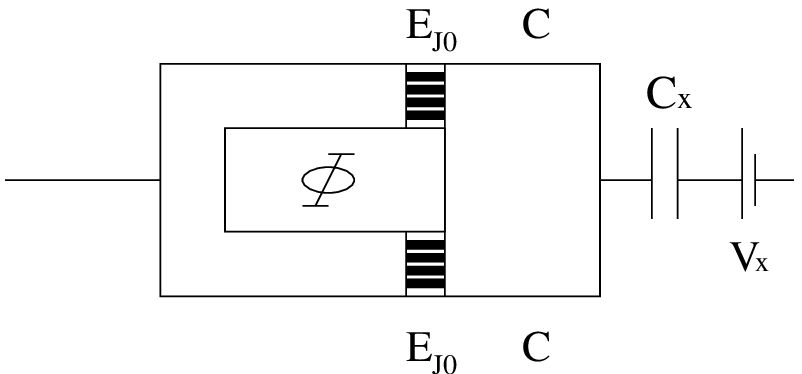}
\end{center}
\caption{ {\bf SQUID with symmetric Josephson junctions} 
}
\label{jo}    \end{figure}

Consider a superconducting electron box formed by a $symmetric$ SQUID( see fig. \ref{jo}), pieced 
by a magnetic flux
$\Phi$ and with an applied gate voltage $V_x$. The device is operated in the charging regime, i.e.,
the Josephson couplings $E_{J0}$
 are much smaller than the charging energy $E_{ch}$. Also a 
temperature much lower than the Josephson coupling is assumed. 
The Hamiltonian for this system
is\cite{vedral,mak,shnirman}
\begin{eqnarray}\label{ha}
H=E_{ch}(\hat n-n_x)^2-E_J(\Phi)\cos(\theta),
\end{eqnarray}
$E_J(\Phi)=2E_{J0}\cos\left(\pi\frac{\Phi}{\Phi_0}\right)$, $E_{ch}$ is the charging energy and $n_x$
can be tuned by the applied voltage $V_x$ through $V_x=2en_x/C_x$( see figure \ref{jo}). The phase difference
across the junction $\chi$ and the cooper pair number $\hat n$ canonical conjugate variables $[\theta,\hat n]=i.$ 
$\Phi_0=h/2e$ is the quantum of flux.
So here $n_x$ and $\Phi$ can be tuned externally. 

If the value of parameter $n_x$ is close to $1/2$, the energy gap between the 
ground state $|0\rangle $ and the first excited state $|1\rangle $ is much smaller than the gaps among any other
states. Thus the basis $|0\rangle , |1\rangle $ approximately make a computational subspace. The transition between
the states in the subspace and the state outside the subspace is small.Through calculation
of the matrix element i.e., $\langle  n_1H|n_2\rangle $( $n_1,n_2=0,1$), 
the matrix form of the Hamiltonian
$H$ in the subspace $|0\rangle ,|1\rangle $ is
\begin{eqnarray}
H_e=-\frac{1}{2}[B_x\sigma_x+B_y\sigma_y+B_z\sigma_z],
\end{eqnarray}
where $(B_x,B_y,B_z)=(E_J(\Phi),-E_J(\Phi),E_{ch}(1-2n_x)$ and $\sigma_{x,y,z}$ are Pauli
matrices. Intutively, we may regard this $H_e$ as the {\it effective} Hamiltonian in the computation
subspace. We can take the rotating operation throhgh the time ecolution property of this effective
Hamiltonian $H_e$.

However, this $H_e$ may not be the best choice in simulating the real time evolution. To reduce
the systematic error,  one can use another form of the effective Hamiltonian which
simulates the time evolution in the computational subspace more precisely.
In this paper, we calculate the matrix element $e^{-iHt}$ by
the purterbation expansion. 
We know the time evolution operator generated by $H$ can be expressed in the form
$e^{-iHt}=\sum_ne^{-iE_n't}|\Psi_n\rangle \langle \Psi_n|$, we still have difficulty in obtaining the precise
time evolution properties for state $|0\rangle $ and $|1\rangle $. Here $E_n'$ and $|\Psi_n\rangle $ are eigenvalue
and eigenstate of $H$, respectively. $|\Psi_n\rangle $ can be expressed in Mathieu function, which is
the infinite summation of trignometric functions. But in the real calculation, we have to use the
truncated Mathieu function. As we will show it latter, after taking the first non-zero perturbation
term into consideration, the error is much smaller than the threshold for 
fault resilent quantum computation\cite{laf}. In the following sections, 
after the we study the single qubit case by the 
perpurbation method, we will investigate the two qubit dynamics, which is the heart of 
the elementary quantum logic
device. We will study both the capasitively coupled SQUIDs and the inductively coupled SQUIDs.
For the former one, we find the effective Hamiltonian in the subspace works exactly. 
For the latter one, we give a new form for the effective Hamiltonian so that it works more
precisely in the computational subspace, especially, the error rate is much lower than the 
threshold for a large, reliable quantum computation.
\section{Time evolution from the perturbation treatment for one qubit}
We first write the Hamiltonian $H$ in the following equivalent form
\begin{eqnarray}
H=E_{ch}\sum_n (n-n_x)^2|n\rangle \langle  n|-\frac{E_J}{2}\sum_{n=0}^{\infty}(|n\rangle \langle  n+1+|n+1\rangle \langle n|).
\end{eqnarray}  
Furthermore, we decompose the Hamiltonian into two parts as
$$
H=H_0+H_c
$$
and
\begin{eqnarray}
H_0=
E_{ch}\sum_n (n-n_x)^2|n\rangle \langle n|-\frac{E_J}{2}(|0\rangle \langle 1|
+|1\rangle \langle 0|)=H_e+E_{ch}\sum_{n> 1}^{\infty}(n-n_x)^2|n\rangle \langle n|,
\end{eqnarray}
\begin{eqnarray}
H_c=\frac{E_J}{2}\sum_{n\not=1}^{\infty}(|n\rangle \langle n+1+|n+1\rangle \langle n|).
\end{eqnarray}
We regard this $H_c$ as the perbation term. $H_0$ part can be solved exactly. The first two eigenvalues($E_0^{(0)}$ and 
$E_1^{(0)}$) and first two
eigenstates($|\phi_0\rangle $ and $|\phi_1\rangle $) are just that of $H_e$. The rest eigenvalues and eigenstates 
are are just that of 
operator $E_{ch}\sum_n (n-n_x)^2|n\rangle \langle n|$. They have already been denoted  as $E_n$ and $|n\rangle $, 
respectively.
If we ignore $H_c$,  this $H_0$ is identitical to $H_e$ in the subspace of $|0\rangle ,|1\rangle $. 
The perturbation mordification comes from $H_c$. We do the perturbation calculation
to the second order of  for the eigenvalues and to the first order to eigenstates here.
(For more exact result, 
one can do it similarly through  higher
order approximation.) 
 We heve the following results:
\begin{eqnarray}
E_0=\frac{E_{ch}(1-2n_x+2n_x^2)-\sqrt{E_{ch}^2(1-2n_x)+E_J^2}}{2}+\Delta E_0,
\end{eqnarray}
\begin{eqnarray}
E_1=\frac{E_{ch}(1-2n_x+2n_x^2)+\sqrt{E_{ch}^2(1-2n_x)+E_J^2}}{2}
+\Delta E_1
\end{eqnarray}
Here $\Delta E_1=\frac{E_J^2}{4E_{ch}(3-2n_x)}$ and $\Delta E_0=\frac{E_J^2}{16E_{ch}n_x}$ 
Take the  perturbation result above into consideration, we obtain 
the time evolution operator  in the computational subspace
\begin{eqnarray}\label{evo}
U(t)=\sum_{n=0}^2 e^{-iE_n t}|\phi_n\rangle \langle \phi_n|.
\end{eqnarray}
Therefore, the new effective Hamiltonian with this timeevolution property is
\begin{eqnarray}\label{nef}
H_e'= E_0|\phi_0\rangle \langle \phi_0|+E_1|\phi_1\rangle \langle \phi_1|
=H_e +\Delta E_0|\phi_0\rangle\langle \phi_0|+\Delta E_1|\phi_1\rangle\langle \phi_1|
\end{eqnarray}
This new effective Hamiltonian gives the evolution operator as Eq. (\ref{evo}).
If $n_x$ is close to $1/2$, the modification in the effective Hamiltonian is insigficant.
Especially, if $n_x=1/2$, $H_e'$ is same with $H_e$, up to a constant term. But they 
may differ 
obviously when $n_x$ is far from $1/2$.  
With this new effective Hamiltonian, if we omit the leakage error, the distance between
the state under real time evolution$(|\psi_r(t)\rangle )$ 
and the state given by evolution of Eq. (\ref{evo}) $(\psi(t))$ is smaller than
$\frac{E_J^4}{E_{ch}^4}$.  
Typically, if $E_J/E_{ch}=0.02$, the error is is between $10^{-8}$ to $10^{-7}$,  smaller
than the threshold for resilent quantum computation, $10^{-6}$ to $10^{-5}$\cite{laf}. However,
if we use the old effective Hamiltonian which is the first order perturbation in the computational
subspace, the error rate is in the magnitude order of $10^{-4}$, larger than the threshold.
However, the error rate can be larger than this threshold if we use the old effective
Hamiltonian $H_e$, provided that $n_x$ is far from $1/2$. 
If the value $E_J/E_{ch}$ is not so small in certain case, we can take a 
higher order purterbation calculation.
\section{two qubit dynamics}
\subsection{capasitive coupling}
For the capasitively coupled SQUID, the interacting
Hamiltonian is
\begin{eqnarray}
H_I=\delta (n_1-n_{x,1})(n_2-n_{x,2}),
\end{eqnarray}
$\delta=2E_{ch}E_{K}/C$.  
Now we regard qubit 1 as the comtroll bit and qubit 2 as the target bit. Our goal is to make a conditional
flip to qubit 2, i.e., if state of qubit 1 is $|0\rangle $, nothing happens to qubit 2; if state of qubit 1 is 
$|1\rangle $, then qubit 2 is flipped.
To make the C-NOT gate, we set $n_{x,1}$ to $0$ and $\Phi_1=\Phi_0/2$ for qubit 1. For qubit 2 we set
the magnetic flux $\Phi_2=\Phi_0$ so that $E_J=0$; and $n_{x,2}=1/2$. Just wait for a period
of 
\begin{eqnarray}
\Delta t=\pi/\delta.
\end{eqnarray}
We obtain the following conditional unitary transformation for qubit 2
\begin{eqnarray}
U=\left(\begin{array}{cc}1&0\\0&1\end{array}\right), 
\end{eqnarray}
if state of qubit 1 is $|0\rangle $; and
\begin{eqnarray}
U=\left(\begin{array}{cc}-i&0\\0&i\end{array}\right), 
\end{eqnarray}
if state of qubit 1 is $|1\rangle $.
This is equivalent to C-NOT gate through Hardmard transformation.
Since $E_J$ is set to 0 here, so the above two qubits operation is exact.
Thus {\it if we use the capasitively coupled SQUIDS, there is no phase shift or leakage error caused by the the states outside the computational
subspace
in the two qubits operation.}
\subsection{Inductively coupled case} 
The C-NOT gate of inductively coupled SQUID system is given in \cite{mak} by using the 
effective Hamiltonian in the computational system. Here we use the time evolution operator
in the whole system and then to calculate the time $t$ at which the fidelity between
time evolution operator and a C-NOT matrix has
the maximum value.

The Hamiltonian for the two interacting SQUIDs is
\begin{eqnarray}
H=H_1+H_2-\frac{1}{E_L}(E_{J1}(\Phi_1)\sin \Theta_1+E_{J2}(\Phi_2)\sin\Theta_2)^2
\end{eqnarray}
We know $\sin\Theta_i= (|n_i\rangle\langle n_i+1|-|n_i+1\rangle\langle n|)/(2i)$.
 We denote
\begin{eqnarray}
H_O=H_0\otimes 1+1\otimes H_0
+\frac{E_{J1}E_{J2}}{2E_{L}}\left[ (|0\rangle\langle 1|-|1\rangle\langle 0|)\otimes
(|0\rangle\langle 1|-|1\rangle\langle 0|)
\right]
\end{eqnarray}
and
$$H_P=H_c\otimes I+I\otimes H_c +
\frac{E_{J1}E_{J2}}{2E_{L}} \left(\sum_{n\not=0} 
|n\rangle\langle n+1|-|n+1\rangle\langle n|\right)\otimes
\left( \sum_{n\not=0} |n\rangle\langle n+1|-|n+1\rangle\langle n|\right) 
$$ 
\begin{eqnarray}
+\frac{E_{J1}^2}{E_L}
\left(\sum_{-\infty}^{+\infty} |n\rangle\langle n+2| +|n+2\rangle\langle n|\right) \otimes 1
+\frac{E_{J2}^2}{E_L}\otimes\left(\sum_{-\infty}^{+\infty} |n\rangle\langle n+2| +|n+2\rangle\langle n|\right) 
.
\end{eqnarray}
Previously, the $H_O$ above was used for the effective Hamiltonian in the computational subspace
for two qubits operation\cite{rev,mak,shnirman}
It is easy to see the total Hamiltonian for the two SQUIDS is
\begin{eqnarray}
H=H_O+H_P.
\end{eqnarray}
We can regard $H_P$ as the perturbation term.
In the two qubit operation, we can always set $n_{x1}=n_{x2}=1/2$. Thus $H_e$ is simplified
to $E_J(\Phi)\left(|0\rangle\langle 1|+|1\rangle\langle 0|\right)$. Under this condition
 we have
\begin{eqnarray}
H_O=U\left(\begin{array}{cccc}
E_{J1}+E_{J2} & 0&0& \frac{E_{J1}E_{J2}}{E_L} \\
0&E_{J1}-E_{J2} & - \frac{E_{J1}E_{J2}}{E_L} &0 \\
0 & -\frac{E_{J1}E_{J2}}{E_L} & E_{J2}-E_{J1} & 0\\
\frac{E_{J1}E_{J2}}{E_L}&0&0 & -E_{J1}-E_{J2} 
\end{array}\right)U^{\dagger}
,\end{eqnarray}
where we have used  the basis of $|00>, |01>,|10> ,|11>$ and 
$U=e^{-\frac{\pi}{4}\sigma_y}\otimes e^{-\frac{\pi}{4}\sigma_y}$. 
Explicitly, $e^{-\frac{\pi}{4}\sigma_y}
=\frac{\sqrt 2}{2}\left(\begin{array}{cc}1 & -1\\1 &1\end{array}\right)$
We can obtain the eigenvalues
and eigenstates of $H_O$ exactly. The four eigenvalues are
$ \pm \sqrt{(E_{J1}+E_{J2})^2+\frac{E_{J1}^2E_{J2}^2}{E_L^2}}$, 
 $\pm \sqrt{(E_{J1}-E_{J2})^2+\frac{E_{J1}^2E_{J2}^2}{E_L^2}}$, corresponding to the following
four eigenstates respectively
$$|\psi_{00}\rangle=U\left(\cos\frac{\theta_1}{2}|00\rangle+\sin \frac{\theta_1}{2}|11\rangle\right)
=\frac{1}{2}\left[\chi_+(|00\rangle+|11\rangle)+\chi_-(|01\rangle+|10\rangle )\right],$$ 
$$|\psi_{01}\rangle=U\left(\cos\frac{\theta_2}{2}|01\rangle+\sin\frac{\theta_2}{2}|10\rangle\right)
=\eta_+(-|00\rangle+|11\rangle)+\eta_-(|01\rangle-|10\rangle),$$ 
$$|\psi_{10}\rangle=U\left(\sin\frac{\theta_2}{2}|01\rangle-\cos\frac{\theta_2}{2}|10\rangle\right)
=\eta_+(-|00\rangle+|11\rangle)-\eta_-(|01\rangle-|10\rangle,$$ 
and $$|\psi_{11}\rangle=U\left(\sin\frac{\theta_1}{2}|00\rangle-\cos\frac{\theta_1}{2}|11\rangle\right)
=\frac{1}{2}\left[\chi_+(|00\rangle+|11\rangle)-\chi_-(|01\rangle+|10\rangle)\right].$$
Here $\chi_{\pm}=\sin\frac{\theta_1}{2}\pm\cos\frac{\theta_2}{2}$, 
$\eta_{\pm}=\sin\frac{\theta_2}{2}\pm\cos\frac{\theta_2}{2}$
, $\cos\theta_1=(E_{J1}+E_{J2})/
\sqrt{(E_{J1}+E_{J2})^2+\left(\frac{E_{J1}E_{J2}}{E_L}\right)^2}$, 
$\sin\theta_1
=E_{J1}E_{J2}/\left(E_L\sqrt{(E_{J1}+E_{J2})^2+\left(\frac{E_{J1}E_{J2}}{E_L}\right)^2}\right)$,  
$\cos\theta_2=E_{J1}-E_{J2}/\sqrt{(E_{J1}-E_{J2})^2+\left(\frac{E_{J1}E_{J2}}{E_L}\right)^2}$ and 
$\sin\theta_2=-E_{J1}E_{J2}/\left(\sqrt{(E_{J1}+E_{J2})^2+\left(\frac{E_{J1}E_{J2}}{E_L}\right)^2}E_L\right)$. 
\\Now we take the first non-zero modification in the perturbation to the 4 states.
The modifications to the 4 energy levels above are
$$\Delta_+,0,0\Delta_-, $$ where
$$
\Delta_{\pm}=\chi_+\Delta E_{11}\pm\chi_-\Delta E_{01}, 
$$ 
 and
\begin{eqnarray}
\Delta E_{11} =\frac{1}{8E_{ch}}
\left[E_{J1}^2+E_{J2}^2
+\frac{1}{2}\left(\frac{E_{J1}E_{J2}}{E_{L}}\right)^2
+\frac{2}{3}\left(\frac{E_{J1}^2+E_{J2}^2}{E_{L}}\right)^2\right]
\end{eqnarray}
\begin{eqnarray}
\Delta E_{01}=\frac{1}{8E_{ch}}
\left[E_{J1}^2+E_{J2}^2-\frac{1}{2}\left(\frac{E_{J1}E_{J2}}{E_{L}}\right)^2
+\frac{2}{3}\left(\frac{E_{J1}^2+E_{J2}^2}{3E_{L}}\right)^2\right]
\end{eqnarray}
Up to an unimportant constant term, the new effective Hamiltonian is
\begin{eqnarray}
H_E=H_O+\Delta_+|\psi_{00}\rangle\langle \psi_{00}|+\Delta_-|\psi_{11}\rangle\langle \psi_{11}|
\end{eqnarray}
Suppose initially we have a state $|\Psi_i\rangle$, after time  $t$, the state in the computational
subspace is $\rho(t)=\Pi e^{-iH t}|\Psi_i\rangle\langle\Psi_i|e^{iHt}\Pi$, 
$\Pi=\sum |kj\rangle\langle kj|$ and $k,j$ can 
take values of 0 and 1. However, if we use the effective Hamiltonian, the state we supposed 
in the computational subsspace is 
$\rho_e(t)=\Pi e^{-iH_E t}|\Psi_i\rangle\langle\Psi_i|e^{iH_Et}\Pi$.  
To estimate the error rate caused by the phase shift error due to the effective Hamiltonian, 
we just calculate distance between $\rho(t)$ and $\rho_e(t)$. Suppose the values of 
$E_{J1}$, $E_{J2}$ are close,
 $E_L$ is not larger than $E_{J1}$ or $E_{J2}$.
After calculation we know, if we use the $H_O$ for the effective Hamiltonian, 
the magnitude order of the distance can be $\frac{E_{J1}^2}{10E_{ch}^2}$. However, if we use the modified
one, $H_E$ for the effective Hamiltonian, the error rate is reduced to the magnitude order of
$\frac{E_{J}^4}{10E_{ch}^4}$. This is to say, if $H_O$ is used as the effective Hamiltonian, 
the error rate is larger than threshold for the
fault resilent quantum computation, $10^{-6}$ to $10^{-5}$, 
provided that  $E_J/E_{ch}$ is larger than $0.01$. However, if we use $H_E$, the one proposed in this 
paper, the error rate is much smaller than the threshold value for the fault resilent quantum computation.

In summary, the properties of the effective Hamiltonian of the Josephson junction system
in the computational subspace is investigated. 
For capasitively coupled Josephson junction system,
 there is
no systematic error due to the effective Hamiltonian in the computational subspace for the two qubits operation.   
If the inductively coupled SQUIDs are used, the effective Hamiltonian in the computational subspace
causes phase shift error. 
But the this effective Hamiltonian can be modified to a more exact form.
 Our new effective Hamiltonian given by the perturbation theory reduces
the systematic error to a range much lower than the threshold of the fault tolerate quantum computation.

{\bf Acknowledgement:} We thank Prof Imai for support. We thank Dr. Huang W.Y for discussions.

\newpage
\newpage
\begin{figure}
\begin{center}
\epsffile{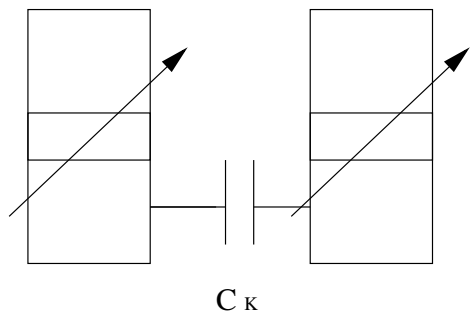}
\end{center}
\caption{ {\bf The capasitively coupled SQUIDs.} 
}
\label{couple1}    \end{figure}
\end{document}